# Wide-Range Probing of Dzyaloshinskii–Moriya Interaction

Duck-Ho Kim,[1] Sang-Cheol Yoo,[1,2] Dae-Yun Kim,[1] Byoung-Chul Min,[2] and Sug-Bong Choe[1†]

[1]Department of Physics and Institute of Applied Physics, Seoul National University, Seoul, 08826, Republic of Korea.

[2]Center for Spintronics, Korea Institute of Science and Technology, Seoul, 02792, Republic of Korea.

[†]Correspondence to: sugbong@snu.ac.kr

**Dzyaloshinskii–Moriya interaction (DMI) in magnetic objects is of enormous interest, because it generates a built-in chirality of magnetic domain walls (DWs) and topologically-protected skyrmions for efficient motion driven by spin–orbit torques. Because of its importance for perspective applications and academic curiosities, many experimental efforts have been devoted to DMI investigation. However, current experimental probing techniques cover only limited ranges of the DMI with specific sample requirements, and there are no versatile techniques covering a wide range of DMI. Here, we present a unique experimental scheme to quantify DMI over a wide range based on the angular dependence of asymmetric DW motion. It can determine DMI even larger than the maximum magnetic field strength, demonstrating that various strengths of DMI can be quantified using a single measurement setup. This scheme provides a standard technique over a wide range of DMI, which is essential to DMI-related emerging fields in nanotechnology.**

**Introduction**

Dzyaloshinskii–Moriya interaction (DMI) is an antisymmetric exchange interaction that occurs at interfaces between ferromagnetic and heavy metal layers with a large spin–orbit coupling[1–3]. In magnetic systems, DMI generates chiral spin textures such as Néel domain walls (DWs)[4-7] and magnetic skyrmions[8–10]. Because these chiral spin textures promise several perspective applications[4,7,8], it is crucial to quantify the strength of DMI for physical exploration toward the origin of DMI and technical optimization of DMI strength in ferromagnetic materials.

To measure DMI strength, several experimental schemes have been proposed[11–15]. Use an optical microscope, Je et al.[11] developed a scheme to estimate DMI strength based on asymmetric DW speed with respect to in-plane magnetic field[11,16,17]. Moon et al.[12] suggested another scheme based on frequency nonreciprocity, which provides a way to measure the DMI constant. Other measurement schemes based on the nonreciprocal propagation of spin waves were also demonstrated using the Brillouin light scattering (BLS)[18–20] and inductive ferromagnetic resonance (FMR)[21] techniques. These techniques are applicable to different measurement ranges of DMI strength with different sample requirements.

The optical microscopy technique[11,16,17] based on asymmetric DW speed provides an easy and simple way to directly measure DMI-induced effective magnetic fields. However, its measurement range is limited by the maximum strength of an external magnetic field, which is limited fundamentally by the narrow available space inside the optical setup. Moreover, application of a large external magnetic field to the optical setup requires sophisticated care to prevent artifacts caused by stray fields from electromagnets as well as mechanical, optical, and thermal artifacts from such strong magnetic fields. In this study, we propose a way to overcome the field strength limit by utilizing angled DWs with respect to the direction of an in-plane magnetic field.

**Results**

***DW energy model for a DW with an angle $\theta$***

The case where a DW is placed with an angle $\theta$ with respect to an in-plane magnetic field is shown in the inset of Fig. 1. The DW energy density $\sigma_{\text{DW}}$ can be expressed as a function of the in-plane magnetic field $H_x$ and magnetization angle $\psi$ from the direction normal to the DW:

$$\sigma_{\text{DW}}(H_x, \psi) = \begin{cases} \sigma_0 + 2\lambda K_{\text{D}} \cos^2 \psi \\ -\pi\lambda M_{\text{S}}[(H_x \cos\theta + H_{\text{DMI}})\cos\psi + H_x \sin\theta \sin\psi] \end{cases}, \quad (1)$$

where $\sigma_0$ is the Bloch-type DW energy density, $\lambda$ is the DW width, $K_{\text{D}}$ is the DW anisotropy energy density, $M_{\text{S}}$ is the saturation magnetization, and $H_{\text{DMI}}$ is the DMI-induced effective magnetic field in the direction normal to the DW. The second term in the equation corresponds to the DW anisotropy energy and the third term corresponds to the Zeeman energy. Note that Eq. (1) is identical to the Stoner–Wohlfarth equation[22] for torque magnetometry with an additional unidirectional bias from $H_{\text{DMI}}$.

For a given $H_x$, the equilibrium angle $\psi_{\text{eq}}$ can be obtained by the minimization condition $\partial \sigma_{\text{DW}} / \partial \psi |_{\psi_{\text{eq}}} = 0$. Moreover, a numerical estimation of Eq. (1) shows that $\sigma_{\text{DW}}$ has a maximum at $H_x = H_0$, where $H_0$ can be obtained from the maximization condition $\partial \sigma_{\text{DW}} / \partial H_x |_{H_0} = 0$. By solving these minimization and maximization conditions simultaneously, one can readily obtain two coupled equations:

$$4K_{\text{D}} \cos\psi_{\text{eq}} \sin\psi_{\text{eq}} - \pi M_{\text{S}} \left[ (H_0 \cos\theta + H_{\text{DMI}}) \sin\psi_{\text{eq}} - H_0 \sin\theta \cos\psi_{\text{eq}} \right] = 0, \quad (2)$$

$$\cos\theta \cos\psi_{eq} + \sin\theta \sin\psi_{eq} = 0. \tag{3}$$

Equation (3) is identical to the relation $\psi_{eq} = \theta \mp \pi/2$, which implies the DW magnetization stays perpendicular to the direction of $H_0$. Replacing $\psi_{eq}$ using of this relation, Eq. (2) can be rewritten as

$$H_0 = (\pm H_K \sin\theta - H_{DMI})\cos\theta, \tag{4}$$

where $H_K$ ($\equiv 4K_D/\pi M_S$) is the DW anisotropy field. The sign of the first term on the right-hand side of Eq. (4) follows that of $H_{DMI}$, i.e., a plus sign for a positive $H_{DMI}$ and a minus sign for a negative $H_{DMI}$. Note that the well-known relation $H_0 = -H_{DMI}$[11,16,17] can be restored in the limit $\theta \to 0$. Figure 1 plots $H_0(\theta)$ obtained from Eq. (4) (red solid line) together with the numerical solution (black symbols) from Eq. (1). The exact conformity verifies the validity of Eq. (4).

Equation (4) delivers the key idea of this study: one can largely reduce $H_0$ by increasing $\theta$. With this scheme, the magnitude of $H_0$ can be adjusted down to a small experimental range $H_{range}$ of the external magnetic field, which enables one to measure a large $H_{DMI}$ without upgrading the electromagnet. For example, by tilting the DWs up to about 80°, one can measure $H_{DMI}$ up to 1 T using an electromagnet with $H_{range} \sim 200$ mT, which is quite comfortable for conventional optical setups[23]. It is also worth noting that this approach enables one to avoid a number of artifacts from large magnetic fields, such as mechanical instability caused by the induced magnetic moment in the optical setup, magnetooptical effect of the objective lens, and large Joule heating caused by the huge current through the electromagnet.

***Verification of θ-dependence in Pt/Co/AlO_x films***

To verify the feasibility, the present scheme is applied to ferromagnetic Pt/Co/AlO$_x$ films, of which $H_{\text{DMI}}$ is slightly smaller than $H_{\text{range}}$. The measurement procedure of $H_0$ follows Ref. 11, except for initially tilted DWs. The tilted DWs were generated using the thermomagnetic writing technique[7,24]. The images on the right-hand side of Fig. 2 show the displacements of the DWs for various $\theta$ values with respect to the direction of $H_x$ (=120 mT) under the application of a fixed out-of-plane magnetic field $H_z$ (=5.5 mT) bias. Each image was obtained by adding several sequential images during the DW displacement with a constant time step (= 500 ms), and thus, each image simultaneously shows several DWs moving from brighter to darker interfaces in time. One can then measure the DW speed $v$ for each image. The plots on the left-hand side of Fig. 2 shows the normalized DW speed $v/v_{\text{min}}$ in the direction normal to the DW as a function of $H_x$, where $v_{\text{min}}$ is the apparent minimum of $v$. It can be seen that $v(H_x)$ is symmetric under inversion with respect to $H_0^*$ as shown in each plot. Here, $H_0^*$ indicates the inversion symmetry axis where $v$ has a minimum. According to Ref. 11, $v$ follows the creep relation $\ln[v(H_x)/v_0] \propto -[\sigma_{\text{DW}}(H_x)]^{1/4}$, where $v_0$ is the characteristic speed. For the case of clear inversion symmetry with a constant $v_0$, the experimental $H_0^*$ exactly matches $H_0$, and thus, we will denote $H_0^*$ by $H_0$ hereafter.

Figure 3a plots the measured $H_0$ with respect to $\theta$. The red solid line shows the best fit by Eq. (4). The good conformity between the data and fitting curve supports again the validity of the equation. The best fitting parameter $H_{\text{DMI}}$ (= −132 ± 3 mT) well matches the experimental value (= −134 ± 6 mT) measured at $\theta = 0$. Moreover, the best-fit parameter of $H_K$ (= −18 ± 5 mT) falls within the range of previous experimental reports[11,25,26]. The value of $H_K$ can be alternatively measured using independent measurements[11,25,26] or estimated using the relation $H_K \cong (4\ln 2 / \pi^2) M_S t_f / \lambda$[27,28], where $t_f$ is thickness of the magnetic layer.

### Application of present scheme to Pt/Co/AlO$_x$ and Pt/Co/MgO films

To mimic the situation that $H_{\text{range}}$ is limited (< 50 mT), the fitting is applied only for the data (box in the plot) with large $\theta$ ($\geq 70°$) as shown in Fig. 3b. The blue solid line indicates the best fit by Eq. (4) with the fixed value of $H_K$ obtained from Fig. 3a. This approach gives the best-fit parameter $H_{\text{DMI}}$ (= $-138 \pm 12$ mT), which matches again previous values within the experimental accuracy. It is therefore demonstrated that the present approach enables one to measure large $H_{\text{DMI}}$ in an experiment with limited $H_{\text{range}}$. Note that the determined $H_{\text{DMI}}$ is larger than twice of $H_{\text{range}}$.

Because the fitting in Fig. 3b is done with a fixed $H_K$, here we examine the effect of the inaccuracy $\delta H_K$ on $H_K$. The blue dotted lines in Fig. 3b are the best fits for the cases with $\delta H_K = \pm 10$ mT. The error $\delta H_{\text{DMI}}$ is found to be slightly smaller than $\delta H_K$, as expected from the relation $\delta H_{\text{DMI}} = \delta H_K \sin\theta$ for $\theta$ in Eq. (4). Because $H_K$ is commonly within the range of a few tens of mT[11,25,26,27,29], $\delta H_K$ may not typically exceed the range of about $\pm 10$ mT, and thus one can confirm that $\delta H_K$-induced error is not significantly large comparison with other experimental errors. Moreover, this error becomes negligible in practical cases because the present approach is designed for determination of large $H_{\text{DMI}}$ ($\gg \delta H_K$) beyond the experimental $H_{\text{range}}$.

Finally, the present scheme is applied to Pt/Co/MgO films, which exhibit DMI larger than $H_{\text{range}}$. Figure 4a shows $v$ as a function of $H_x$ for $\theta = 0$. This plot clearly shows that the inversion symmetry axis $H_0$ is beyond the experimental $H_{\text{range}}$ (i.e., $H_0 \gg 200$ mT), and thus conventional optical schemes cannot quantify $H_{\text{DMI}}$. However, by applying the present scheme, Fig. 4b shows the measured $H_0$ with respect to $\theta$ for a large range of $\theta$ ($\geq 80°$). The black box in the figure indicates the measurable window for $H_{\text{range}}$ of the present setup. The best fit (blue solid line) of $H_\text{K}$ (= −30 ± 5 mT) indicates that $H_\text{DMI}$ = −483 ± 10 mT, which is more than twice larger than $H_\text{range}$. The sign and magnitude are confirmed to match previously reported results[21]. The blue dotted lines in Fig. 4(b) are the best fits for the cases with $\delta H_\text{K}$ = ±10 mT, and thus it is clearly demonstrated that the error becomes negligible in this case.

**Discussion**

Additional asymmetry from chiral damping[30] or asymmetric DW width variation[27] may cause a shift $\delta H_0$ in $H_0$. However, because the asymmetric slope in $v$ caused by these phenomena appears only during chirality variation occurring within the range of $\pm H_\text{K}$, $|\delta H_0|$ is essentially smaller than $|H_\text{K}|$. Therefore, $\delta H_0$-induced errors are negligible again in practical cases for large $H_\text{DMI}$ determination.

In conclusion, we proposed a scheme to quantify $H_\text{DMI}$ over a wide range by overcoming the limitation from $H_\text{range}$. By measuring the angular dependence of asymmetric DW motion, we clearly find that the $H_0$ has the clearly angle dependence, in a robust manner quantifying the large DMI with large $\theta$. The feasibility of the present scheme is demonstrated experimentally for various strengths of DMI using ferromagnetic Pt/Co/AlO$_x$ and Pt/Co/MgO films. The errors caused by additional asymmetry and inaccuracy of $H_\text{K}$ were found to be negligible in practical cases for large $H_\text{DMI}$ determination. The present scheme enhances the experimental range of optical measurement techniques without upgrading electromagnets. Our finding opens a novel way to easy and straightforward manner to explore materials and systems for large DMI, and thus surmounts the key obstacle to design emerging devices with tailoring the DMI for topological stability and efficient manipulation as required for next-generation nanotechnology.

**Figure Captions**

Figure 1. **Plot of $H_0$ as a function of $\theta$.** Black symbols are obtained by the numerical solution from Eq. (1) and red solid line is calculated by Eq. (4). The inset is schematic illustration for measurement.

Figure 2. **Plot of $v/v_{\min}$ with respect to $H_x$ fixed $H_z$ and under various $\theta$ in Pt/Co/AlO$_x$ film**. **a**, 0° **b**, 30° **c**, 60°, and **d**, 90°. Displacement driven by a magnetic field for DWs at each angle.

Figure 3. **Plot of the measured $H_0$ with respect to $\theta$ in Pt/Co/AlO$_x$ film. a**, Collected data with large $\theta$ ranges from 0° to 90°. The red solid line shows the best fit by Eq. (4). **b**, Collected data with small $\theta$ ranges from 70° to 90°. The blue solid line exhibits the best fit by Eq. (4) with the fixed value of $H_K$. The blue dotted lines in Fig. 3(b) are the best fits for the cases with $\delta H_K = \pm 10$ mT, respectively.

Figure 4. **DMI determination in Pt/Co/MgO film. a**, Plot of the measured $v$ with respect to $H_x$ fixed $H_z$ =20 mT at $\theta$ =0. **b**, Plot of the measured $H_0$ with respect to $\theta$ for the $\theta$ ranges from 80 to 90°. The blue solid line exhibits the best fit by Eq. (4) and the blue dotted lines in Fig. 4(b) are the best fits for the cases with $\delta H_K = \pm 10$ mT, respectively.

**Methods**

Methods and any associated references are available in the online version of the paper.

**Acknowledgements**

This work was supported by a National Research Foundations of Korea (NRF) grant that was funded by the Ministry of Science, ICT and Future Planning of Korea (MSIP) (2015R1A2A1A05001698 and 2015M3D1A1070465). D.-H.K. was supported by a grant funded by the Korean Magnetics Society. B.-C.M. was supported by the KIST institutional program and Pioneer Research Center Program of MSIP/NRF (2011-0027905).

**Author contributions**

D.-H.K. planned and designed the experiment and S.-B.C. supervised the study. D.-H.K. and D.-Y.K. carried out the measurement. S.-C.Y. and B.-C.M. prepared the samples. S.-B.C. and D.-H.K. performed the analysis and wrote the manuscript. All authors discussed the results and commented on the manuscript.

**Additional information**

Correspondence and request for materials should be addressed to S.-B.C.

**Competing financial interests**

The authors declare no competing financial interests.

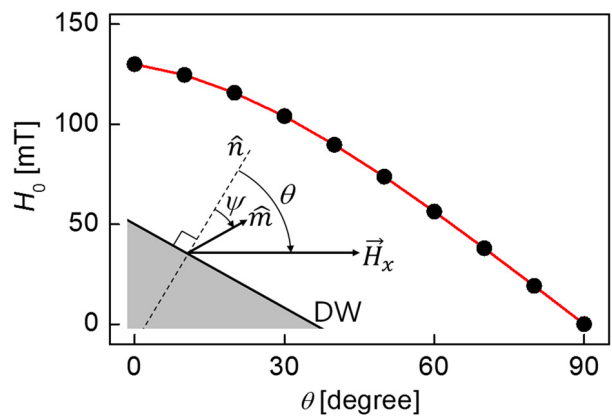

Figure 1

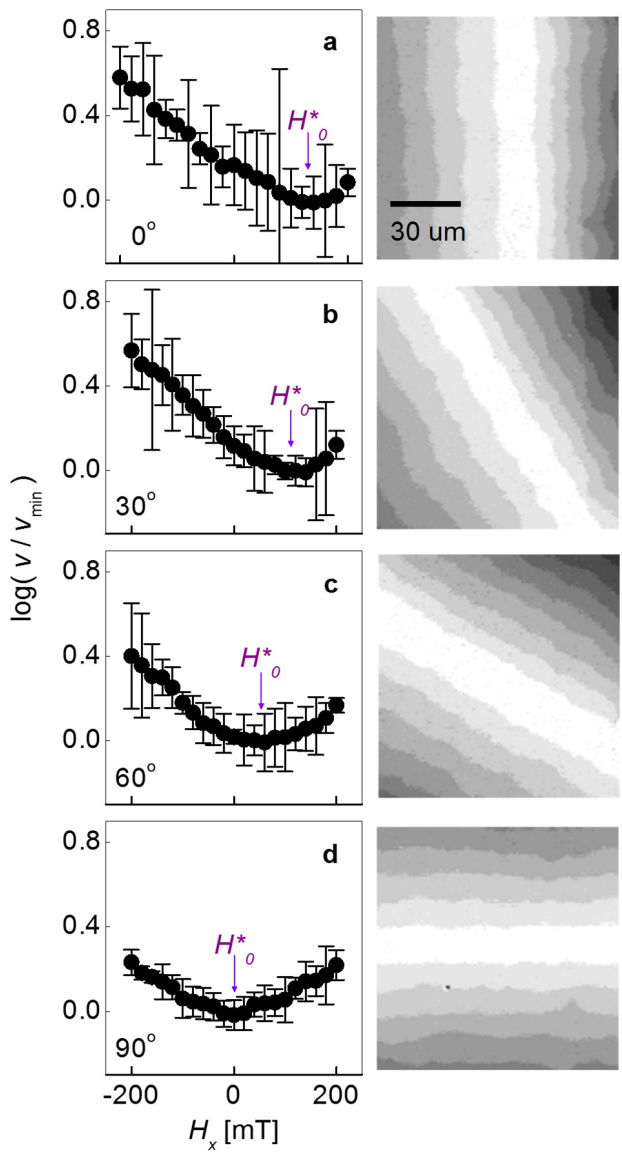

Figure 2

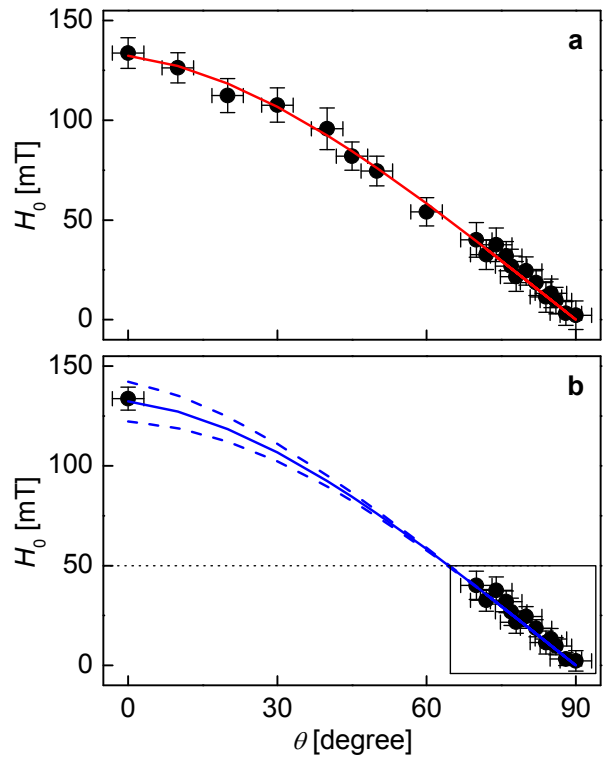

Figure 3

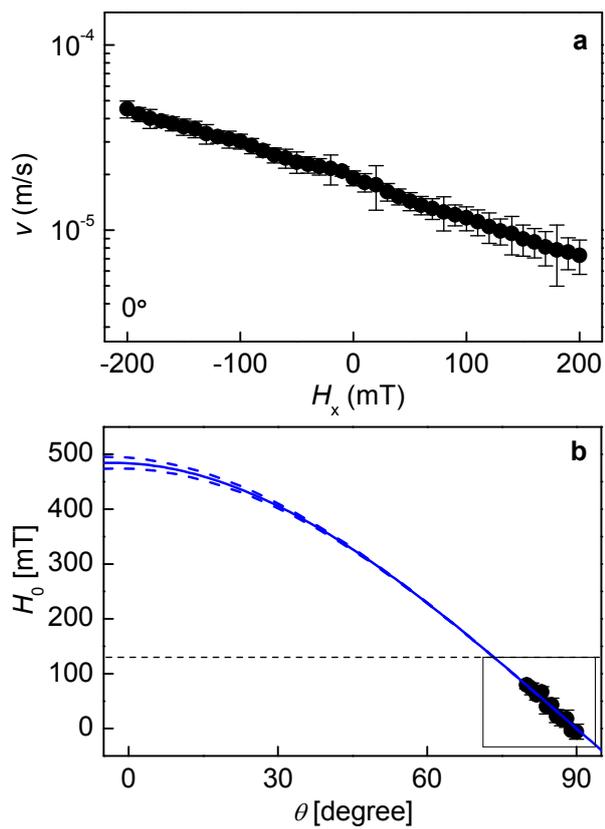

Figure 4